\documentclass[10.5pt,compsoc]{tst}
\usepackage{graphicx}
\usepackage{footmisc}
\usepackage{subfigure}
\usepackage{url}
\usepackage{multirow}
\usepackage[noadjust]{cite}
\usepackage{amsmath,amsthm}
\usepackage{amssymb,amsfonts}
\usepackage{booktabs}
\usepackage{color}
\usepackage{ccaption}
\usepackage{booktabs}
\usepackage{float}
\usepackage{fancyhdr}
\usepackage{caption}
\usepackage{xcolor,stfloats}
\usepackage{comment}
\graphicspath{{figures/}}
\usepackage{cuted}  
\usepackage{epstopdf}
\usepackage{cite}
\usepackage{makecell}  
\usepackage{booktabs}  
\usepackage{tabularx}   
\usepackage{graphicx}

\headevenname{\normalsize{\textbf{\emph{}} }}%
\headoddname{{\sf First author:}\quad {\textbf{\emph{}}}}%

\newtheoremstyle{mystyle}{0pt}{0pt}{\normalfont}{1em}{\bf}{}{1em}{}
\theoremstyle{mystyle}
\renewcommand\figurename{Fig.~}

\newcommand{\nop}[1]{}

\makeatletter
\renewcommand{\@biblabel}[1]{[#1]\hfill}
\makeatother

\begin{document}

\thispagestyle{empty}

\hyphenpenalty=50000

\makeatletter
\newcommand\mysmall{\@setfontsize\mysmall{7}{9.5}}

\newenvironment{tablehere}
  {\def\@captype{table}}
  {}
\newenvironment{figurehere}
  {\def\@captype{figure}}
  {}

\thispagestyle{plain}%
\thispagestyle{empty}%

\let\temp\footnote
\renewcommand \footnote[1]{\temp{\normalsize #1}}
{}
\vspace*{-40pt}

\vskip .2mm

\vskip .2mm

\vskip .2mm\noindent

\begin{strip}
{\center
{\LARGE\textbf{
Hiding Functions within Functions: Steganography by Implicit Neural Representations
}}
\vskip 9mm}

{\center {\sf 
Peng Luo, Jia Liu$^*$, Yan Ke, Qian Dang, Minqing Zhang and Dejun Mu
}
\vskip 5mm}

\centering{
\begin{tabular}{p{160mm}}

{\normalsize
\linespread{1.6667} %
\noindent
\bf{Abstract:} {\sf
Deep steganography utilizes the powerful capabilities of deep neural networks to embed and extract messages, but its reliance on an additional message extractor limits its practical use due to the added suspicion it can raise from steganalyzers. To address this problem, we propose StegaINR, which utilizes Implicit Neural Representation (INR) to implement steganography. StegaINR embeds a secret function into a stego function, which serves as both the message extractor and the stego medium for secure transmission on a public channel. Recipients only need to use a shared key to recover the secret function from the stego function, allowing them to obtain the secret message. Our approach employs continuous functions, enabling it to handle various types of messages. To our knowledge, this is the first work to introduce INR into steganography. We performed evaluations on image, climate data and NeRF synthetic dataset to test our method in different deployment contexts.
}
}
\end{tabular}
}
\vskip 6mm

\vskip -3mm
\small\end{strip}

\thispagestyle{plain}%
\thispagestyle{empty}%
\makeatother
\pagestyle{tstheadings}

\begin{figure}[b]
\vskip -6mm
\begin{tabular}{p{44mm}}
\toprule\\
\end{tabular}
\vskip -4.5mm
\noindent
\setlength{\tabcolsep}{1pt}
\begin{tabular}{p{1.5mm}p{79.5mm}}
&
\\
$\bullet$& Peng Luo is with the School of Cybersecurity, Northwestern Polytechnical University, Xi'an 710072, China and the School of Cryptography Engineering, Engineering University of PAP, Xi'an 710086, China. E-mail: lp\_nwpu@mail.nwpu.edu.cn \\
$\bullet$& Jia Liu, Yan Ke, Qian Dang, Minqing Zhang are with the School of Cryptography Engineering, Engineering University of PAP, Xi'an 710086, China. Email: liujia1022@gmail.com; 15114873390@163.com; dangqian729@163.com; api\_zmq@ 126.com. \\
$\bullet$& Dejun Mu  are with the school of cybersecurity, northwestern polytechnical university, Xi'an 710072, China. Email: mudejun@mail.nwpu.edu.cn. \\
$\sf{*}$&
To whom correspondence should be addressed. \\

\end{tabular}
\end{figure}\large

\vspace{3.5mm}
\section{Introduction}
\label{s:introduction}
\noindent
Steganography is a technique used for secret communication, where confidential information is hidden within a seemingly innocuous cover data and transmitted through a public channel\textsuperscript{\cite{SteganographyinDigital}}. The cover data can take various forms, such as digital text, audio, images, videos, or 3D data. Traditional steganographic methods primarily focus on designing algorithms respectively to subtly modify or select the cover media in order to embed the secret information\textsuperscript{\cite{Coverlessimagesteganography,Minimizingadditivedistortion}}. However, it is challenging to achieve a good balance between concealment\textsuperscript{\cite{Structuraldesignof,Deepresidualnetworkforsteganalysisofdigitalimages,dong2024high}} and embedding capacity.
\begin{figure}[h]\centering
	\includegraphics[width=0.8\linewidth]{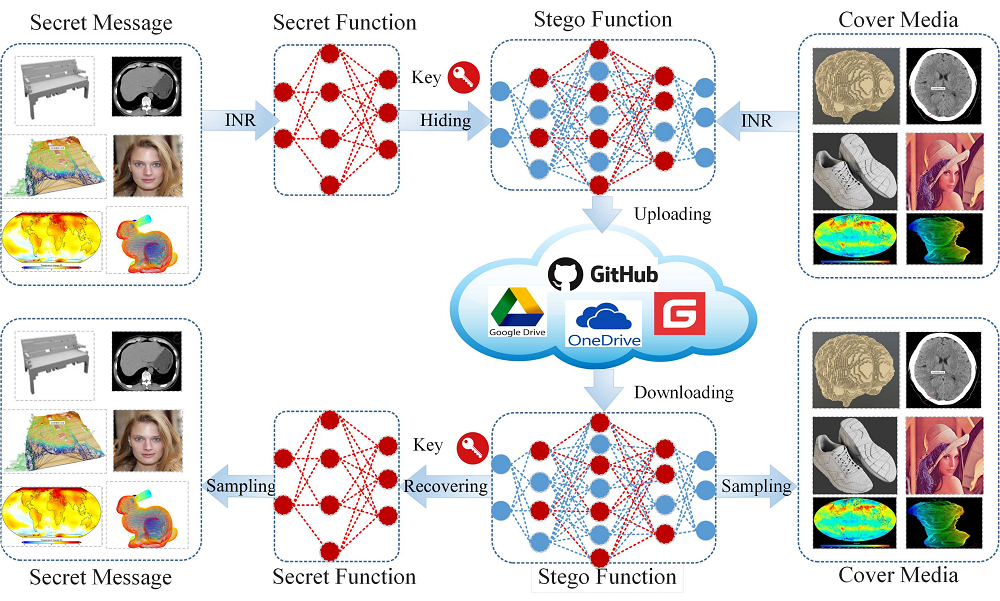}
	\caption{INR-based Steganography: Embedding a Secret Function into a Stego Function. StegaINR enables secure covert communication by embedding and recovering multimodal data via implicit neural representations.}
    \label{FIG_1}
\end{figure}
Recently, deep neural network-based steganography schemes have been proposed, which generally exhibit superior performance in comparison to traditional approaches. In deep steganography, the sender of the message employs an encoding network to embed confidential information into the cover medium, while the receiver of the message makes use of a message decoder to recover the secret information from the cover medium. However, deploying these deep steganography methods in practical applications poses two critical issues. Firstly, these methods necessitate the transmission of a large decoder to the message recipient, but it is insecure to transmit the secret decoder for another transmission. Secondly, a more concealed and serious problem arises when the decoder is present at the recipient’s end, which may arouse suspicion from analysts regarding the steganographic behavior. 
One possible solution is to treat the message decoder as ordinary binary data and hide it in a popular cover data such as an image or video using existing steganography schemes. However, due to the relatively large size of the message decoder, traditional steganography schemes require a significant amount of cover data to hide secret information, which imposes a significant communication burden. Another solution\textsuperscript{\cite{SteganographyofSteganographicNetworks,TowardsDeepNetworkSteganography}} is to embed the decoder network into another deep neural network that performs other tasks, such as image classification or segmentation. However, this stego network 
is even larger than the message extractor and also causes a communication burden. In addition to the concealed transmission of the message decoder, both of these solutions require the transmission of the stego data.

To address the covert issue of message decoders, we propose a novel steganographic scheme based on INR, which parameterizes the signal as a continuous function. In our scheme, as shown in Fig. \ref{FIG_1}, (1) we first represent the secret message as a continuous secret function, (2) by using a shared key, we extend the secret function (network) to construct a new network structure, (3) with fixed parameters of the secret function, we parameterize cover data using the extended function to obtain a stego function. The stego function serves as both an extractor and stego data, enabling secure transmission over a public channel. (4) The recipient only needs to utilize the shared key to recover the secret message from the stego function. Other users without the key can only obtain the cover data from the stego function. Our proposed INR based steganographic technique has the following advantages:

\textit{\textbf{Security}}: In our scheme, the stego function can be seen as both a message extractor and a stego data. The stego function's ability to implicitly represent the cover data conceals its message extraction capability. This not only addresses the security issue of transmitting message decoders over a public channel but also maintains the covertness of the stego function at the receiving end, thereby protecting the steganographer's behavior.

 \textit{\textbf{Capacity}}: Since INR parameterize the secret message as a continuous function, the memory required for parameterized signals is independent of spatial resolution but only dependent on the complexity of the underlying signal. This allows us to achieve high capacity steganography using continuous functions.

\textit{\textbf{Communication Cost}}: {In the actual covert communication process, each transmitted secret message is different. In the scheme proposed in this paper, the secret function is INR of the secret message, and transmitting the secret function is essentially transmitting the secret message. In our method, the extended secret function (stego function) itself serves as the cover data. We only need to transmit the stego function over a public channel to achieve covert communication. Furthermore, implicit representations of continuous functions typically have a smaller size. Compared to \cite{SteganographyofSteganographicNetworks,TowardsDeepNetworkSteganography}}, this significantly reduces the communication cost.

\textit{\textbf{Versatility}}: By transforming secret message data into a unified data format, i.e., a function, our simple method provides a new uniform implementation framework for data hiding, which can be applied to various data types, such as images, 3D models, weather data, and others.

\textit{\textbf{Efficiency}}: Unlike \cite{SteganographyofSteganographicNetworks,TowardsDeepNetworkSteganography}, our approach only requires parameterization on a single data sample, without the need for training on large-scale datasets for other tasks. Furthermore, training implicit representations requires transforming grid data into coordinate and feature-based representation. This representation enables us to randomly sample a subset of elements instead of representing all data, for example, selecting a certain number of pixels from the entire image. Training on subsets eliminates dependence on data resolution, reduces memory consumption, and improves efficiency. 

In particular, the recent development of neural implicit representation technology has made it a trend to widely distribute visual data by sharing implicit representation model weights. An example of this is the NeRF\textsuperscript{\cite{Nerf}} for 3D representation. We believe that in the future, people will share various multimedia data contents that they capture as function formats, just as they currently share 2D images and videos online. In this context, we are also interested in the following research questions: (1) It is common to inject information into 2D images or 3D content for steganography or ownership recognition. But when people share information through implicit representation functions, can we achieve information hiding? (2) How to define steganographic capacity when implementing embedding operations in network with implicit representation? 

StegaINR presents the first exploration of hiding information in INR models. It not only solves the problem of secure transmission of deep steganographic models but also systematically addresses the above two questions. Our contributions can be summarized as follows:

\begin{itemize}
    \item We propose the new problem of implicit representation steganography and strive to embed customizable, imperceptible, and recoverable secret information in continuous functions.
    \item We solve the problem of secure transmission of deep decoding networks and high communication cost by disguising the secret message as a camouflage function.
    \item We propose a key-based stego function construction strategy, which utilizes the structure of the stego function constructed from the shared secret information and trains the parameters of the stego function through a mask-based partial optimization strategy.
    \item We empirically verify the proposed framework on diverse datasets with different data types, such as images and climate data. We achieve high accurate recovery while ensuring embedding capacity.
\end{itemize}
\section{Related Work}\label{sec2}

\subsection{Deep Steganography}

Deep steganography refers to a technique that employs deep neural networks to implement covert communication. These methods are mainly divided into two categories. One category is based on encoder-decoder networks. They utilize the powerful encoding ability of deep neural networks to encode the cover data and message into the stego data. Baluja\textsuperscript{\cite{baluja2020hiding}} first proposed the use of deep neural networks as encoders to hide secret images into other images, and they also trained a decoder to recover the message. Zhu et al.\textsuperscript{\cite{Hidden:Hidingdatawithdeepnetworks}} introduced a noise layer into this framework, allowing the decoder to achieve robust message recovery. Jing et al.\textsuperscript{\cite{jing2021hinet}} unified the encoder and decoder into a reversible network to achieve deep steganography. They also implemented multi-image embedding to achieve high-capacity steganography\textsuperscript{\cite{DeepMIH}}. Xu et al. proposed a flow-based robust reversible image steganography framework\textsuperscript{\cite{Robustinvertibleimagesteganography}}. Recently, a special method for steganography through fixed neural networks\textsuperscript{\cite{FIXEDNEURALNETWORKSTEGANOGRAPHY}} has been proposed, which adds controllable noise to the carrier to avoid training the message extraction network. These methods require an existing natural and realistic cover to construct stego data. 

Another category of methods uses deep generative models, such as GANs, diffusion models, flow models, and autoregressive encoders to construct cover data, stego data or modification policies. Volkhonskiy et al.\textsuperscript{\cite{Steganographicgenerativeadversarialnetworks}} first used GANs to generate an original cover image to counter steganalysis. Liu et al.\textsuperscript{\cite{Coverlessinformationhidingbased}} use a category-controlled GAN to establish a mapping relationship between the message and generated images to hide information. Similar to \cite{Steganographicgenerativeadversarialnetworks}, Hu et al. establishe a mapping relationship between messages and noise images and then uses the GAN to generate a stego image directly\textsuperscript{\cite{Anovelimagesteganographmethod}}. Zhang et al. regarded the generator as an image sampler and extracted the message using a fixed message decoder called the Carden Grille\textsuperscript{\cite{GenerativeSteganographybySampling}}. Tang et al. used GANs to generate distortion cost functions to implement cover modification steganography\textsuperscript{\cite{Automaticsteganographicdistortionlearning,Spatialimagesteganography}}. Chen et al. directly generates stego using variational autoencoders\textsuperscript{\cite{Auto-encodingvariationalbayes,Embeddingwatermarksintodeepneuralnetworks}} (VAE) and flow model-based generators\textsuperscript{\cite{Glow:Generativeflowwithinvertible}}\textsuperscript{\cite{WhenProvablySecureSteganographyMeetsGenerativeModels}}. Yang et al. proposed a new framework for generative steganography using autoregressive models, which uses the pixel CNN+ to generate hidden steganography\textsuperscript{\cite{ProvablySecureGenerativeSteganography}}. Zhang et al.\textsuperscript{\cite{Pixel-Stega}} also used autoregressive models to design steganography schemes and implemented pixel-level information hiding using Pixel CNN+ and arithmetic coding. Liu et al. decoupled images into structural and textural features using autoencoders, enabling stable feature extraction from structural features to improve the accuracy of secret message extraction\textsuperscript{\cite{ImageDisentanglementAutoencoderforSteganographyWithoutEmbedding}}. Recently, there have also been schemes using diffusion models\textsuperscript{\cite{DenoisingDiffusionProbabilisticModels,DenoisingDiffusionImplicitModels}} to construct steganography. Wei et al. embed the message using modified DC coefficients of the diffusion noise, and then a reverse diffusion model restores the noise to a natural image\textsuperscript{\cite{GenerativeSteganographyDiffusion}}. Finally, the message is extracted using a diffusion process and IDCT transform. Similar to  Hu et al.\textsuperscript{\cite{Anovelimagesteganographmethod}}, Kim et al.\textsuperscript{\cite{Diffusion-Stego}} establishes a mapping relationship between messages and noise images and then used a diffusion model to generate stego data. Yu et al. uses the image conversion ability of the diffusion model and the robustness of noise to adopt a stable diffusion conditional model to implement an image steganography\textsuperscript{\cite{CRoSS:DiffusionModel}}.

In order to achieve high-quality coding of cover data and decoding of messages, these deep learning-based steganography schemes usually require a lot of computing resources to obtain powerful encoders (generator) and decoders. These networks are usually trained on grid-like data such as images, making the size of these encoders and decoders directly related to the resolution of the underlying grid data. Except for \cite{GenerativeSteganographybySampling} and \cite{FIXEDNEURALNETWORKSTEGANOGRAPHY}, all the above methods need to transfer the message extractor with a large amount of data to the message receiver, which not only increases the communication burden but also increases the risk of exposing steganographic behavior. To solve this problem, Li et al.\textsuperscript{\cite{SteganographyofSteganographicNetworks,TowardsDeepNetworkSteganography}} disguise a steganography network (called a secret DNN model) by ordinary deep neural networks (called stego DNN) with image classification or segmentation tasks. In \cite{SteganographyofSteganographicNetworks}, they disguised the entire secret message extractor by selecting and adjusting some of the filters to retain their functionality in the secret task. In \cite{TowardsDeepNetworkSteganography}, they proposed a gradient-based filter integration scheme to insert interference filters into important positions in the secret DNN model to form a stego DNN model. Although these two methods ensure the security of the message decoder, they still do not avoid the problem of high communication burden. These two methods use image processing behavior (classification or segmentation) to conceal steganography behavior. They only provide a concealment strategy for message extractors in steganographic schemes. In this paper, we present a complete steganographic framework for multimedia data and other data, where we represent the message as a continuous function and embed this function into another continuous function. This not only provides security for message extractors but also eliminates the need for additional transmission of stego data, as the stego function itself acts as both a stego and extractor. Additionally, the resolution of the neural implicit representation is decoupled from the training data, and implicit model representations typically have smaller data sizes, thus avoiding high-load communication.

\subsection{Model Hiding}
With the widespread use of model data in the cloud, researchers have also paid extensive attention to information hiding in model data\textsuperscript{\cite{qin2025advances}}. Model hiding can be mainly divided into two aspects: \textit{Neural Network Steganography} and \textit{Neural Network Watermarking}. Neural network watermarking can be classified into three types: white-box watermarking, black-box watermarking, and no-box watermarking. In white-box watermarking, The verifier can access the internal structure of the network, including the weights, to verify the copyright. Uchida et al. proposed a method to embed a watermark consisting of 0s and 1s into the weights of the network through regularization, and the watermark can be extracted using a specific algorithm to verify the copyright\textsuperscript{\cite{Embeddingwatermarksintodeepneuralnetworks}}. Black-box watermarking is suitable when the verifier cannot access the internal structure of the network and can only use remote API calls. A common approach is to select specific samples as triggers and fine-tune the network to fit these triggers for copyright verification\textsuperscript{\cite{Turningyourweaknessintoastrength}}. No-box watermarking is mainly used for copyright verification of generative networks\textsuperscript{\cite{Watermarkingneuralnetworks}}. The strategy is to train the network to generate images that contain the watermark, and the verifier can directly verify the copyright from the generated images. 

Some scholars have proposed the model steganography where the model data is used as stego data for hidden information. Yang et al.\textsuperscript{\cite{Ageneralsteganographic}} embedded secret data into the convolutional layers of a given neural network (the cover network) during the network training process. Matrix multiplication is used to encode the parameters of the convolutional layers for data extraction. Chen \textsuperscript{\cite{HidingImagesinDeepProbabilisticModels}} used a deep neural network (DNN) to establish a probability density model of cover images and hide a secret image at a specific location of the learned distribution. They used SingGAN \textsuperscript{\cite{SingGAN}} to learn the distribution of patches in a cover image. During the patch distribution learning process, a deterministic mapping is fitted from a fixed set of noise maps (generated by embedding keys) to hide the secret image. Yang et al. proposed a multi-source data hiding scheme for neural networks, where multiple senders can simultaneously send different secret data to a receiver using the same neural network\textsuperscript{\cite{Multi-source}}. They achieved data embedding during the training process of the neural network, replacing post-training modifications to the network. 

These methods treat neural networks as a special type of data asset, similar to images or video in traditional watermarking or steganography. Unlike the previous model watermarking and steganography, in this paper, we view neural networks as another representation of the secret data. From this perspective, we transform the secret message into a unified data format, which is a function, providing a consistent implementation framework for data hiding in images, 3D models, weather data, and other types of data.

\subsection{Information Hiding For NeRF}
Recently, a special type of neural implicit representation for 3D data called Neural Radiance Fields (NeRF)\textsuperscript{\cite{Nerf}} has received extensive research attention. Li et al.\textsuperscript{\cite{StegaNeRF}} proposed StegaNeRF, a method to embed hidden information in NeRF rendering. They design an optimization framework that first trains a regular NeRF model and then performs a second training to achieve the rendering of the secret image while ensuring the original visual quality. Similarly, Luo et al. proposed an anti-distortion rendering scheme by replacing the original color representation in NeRF with a watermarked color representation, ensuring stable extraction of the watermark information from the 2D images rendered by NeRF.\textsuperscript{\cite{CopyRNeRF}}. Chen et al.\textsuperscript{\cite{WatermarkingforNeuralRadianceField}} utilized the synthesized novel views by NeRF as an important basis for copyright verification and trained a watermark verification model using a parameterized approach. Similar to \cite{WatermarkingforNeuralRadianceField}, Dong et al. \textsuperscript{\cite{SteganographyforNeuralRadianceFieldsbyBackdooring}} applied this technique to steganography and developed a NeRf-based steganography scheme. These approaches treat the neural radiance field as a new-view image generator, embedding secret information in the images from the new views. Therefore, an additional decoder is still required for message extraction. In this paper, we embed the implicit representation function of the secret message into the implicit representation function of the cover data. We can directly extract the secret message from the stego function using the shared key between the sender and receiver, avoiding the need for the message extractor.
\section{Representing Data As Functions}
In this section, we will provide a comprehensive overview of INR, with a focus on using images as an illustrative example.

\subsection{Representing an Image with Function}
Suppose we have an image where $I[x;y]$ corresponds to the RGB value at pixel position $(x;y)$. We want to represent this image as a function $f: {{R}^{2}} \rightarrow {{R}^{3}}$, where $f(x;y) = (r;g;b)$ returns the RGB value at pixel position $(x;y)$. To achieve this, we parameterize the function $f_{\theta}$ with a multi-layer perceptron (MLP) with weighted parameters $\theta$, which is often referred to as an $INR$. We can then learn this representation by minimizing the following loss function\textsuperscript{\cite{GenerativeModelsasDistributions}} as described in Eq. (\ref{eq1}).
\begin{equation}
    {{L}_{I}} = \min_{\theta} \sum_{x,y} \left\| f_{\theta}(x,y) - I[x,y] \right\|_2^2   \label{eq1}
\end{equation}

where the sum is over all pixel positions. It is worth noting that the representation $f_{\theta}$ does not depend on the number of pixels. Therefore, unlike most image representations, the representation $f_{\theta}$ is resolution-independent \textsuperscript{\cite{Occupancynetworks,Implicitneuralrepresentationswithperiodicactivationfunctions}}.

\subsection{Representing General Data With Functions}
This example for images can be generalized to more general data. Let $x \in  X$ represent the coordinates and $y \in  Y$ represent the features, and suppose we have a set of data points with pairs of coordinates and features $\{ (x_i, y_i) \}_{i=1}^n$. For example, for an image, $x = (x;y)$ corresponds to the pixel positions, and $y = (r;g;b)$ corresponds to the RGB values, and $\{ (x_i, y_i) \}_{i=1}^n$ corresponds to the set of all pixel positions and RGB values. Given a set of coordinates and their corresponding features, we can learn a function $f_{\theta}: X \rightarrow Y$ to represent this data by minimizing the following loss function as described in Eq. (\ref{eq2}).
\begin{equation}
L = \min \sum_{i=1}^n \left\| f_{\theta}(x_i) - y_i \right\|_2^2                 \label{eq2}
\end{equation}

A key characteristic of these representations is that they scale with the complexity of the signal rather than the scale of the signal. In fact, for images, the memory required to store the data grows quadratically with the resolution, while for voxel grids, it grows cubically. In contrast, for function representations, the memory requirements are directly related to the complexity of the signal: to represent a more complex signal, we need to increase the capacity of the function $\theta$, for example by increasing the depth of the neural network.
\begin{figure*}[t]\centering
	\includegraphics[width=\linewidth]{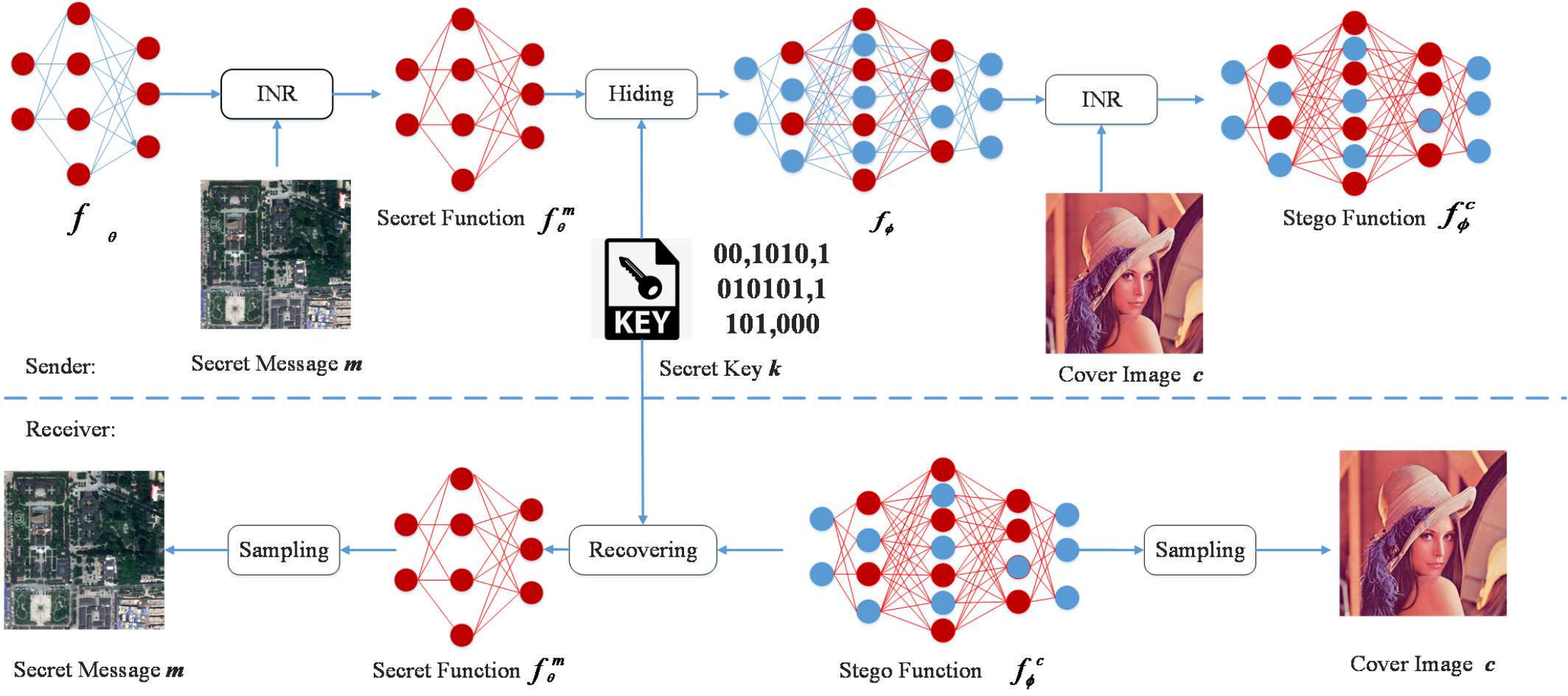}
	\caption{Steganography Framework based on Implicit Neural Representations.}
    \label{FIG_2}
\end{figure*}

Recent research has shown that learning function representations by minimizing Eq. (\ref{eq1}) results in a bias towards low-frequency functions\textsuperscript{\cite{Nerf,Implicitneuralrepresentationswithperiodicactivationfunctions}}. While several methods have been proposed to mitigate this issue, we use the Random Fourier Features (RFF) encoding introduced by Tancik \textsuperscript{\cite{Fourierfeaturesletnetworkslearn}} as it does not bias towards axial changes \textsuperscript{\cite{Nerf}} and does not require specialized initialization \textsuperscript{\cite{Implicitneuralrepresentationswithperiodicactivationfunctions}}. Specifically, for a given coordinate $x \in {{R}^{d}}$, the encoding function $\gamma: {{R}^{d}} \rightarrow  {{R}^{2m}} $ is defined as follows:

\begin{equation}
\gamma(x) = \left( \begin{matrix} \cos(2\pi Bx) \\ \sin(2\pi Bx) \end{matrix} \right)      \label{eq3}
\end{equation}
where $B$ is a random matrix of size $m \times d$ (possibly learnable) whose elements are typically sampled from $N(0,\sigma^2)$. The number of frequencies $m$ and the variance of the elements of matrix $B$, $\sigma^2$, are hyperparameters. To learn high-frequency functions, we simply encode $x$ before passing it through the MLP $f_{\theta}$, i.e., $f_{\theta}(\gamma(x))$, and then minimize Eq. \eqref{eq1}. Using an MLP with ReLU activation to learn the function representation of an image fails to capture high-frequency details, while using RFF encoding followed by a ReLU MLP allows accurate reproduction of the image.

\section{Problem Formulation}

\subsection{Function Hiding}
For simplicity, let's use the implicit representation of images as an example to describe our method, as shown in Fig. \ref{FIG_2}. Assuming the secret message to be transmitted is denoted as $m$, for the sender, the following three steps need to be completed sequentially to achieve steganography:

\begin{equation}
    {{f}_{\theta }^{m}}=IN{{R}_{p}}(m,{{f}_{\theta}})  \label{eq4}
\end{equation}
\begin{equation}
    {{f}_{\phi}}=Hide({{f}_{\theta }^{m}},k)   \label{eq5}
\end{equation}
\begin{equation}
    {{f}_{\phi }^{c}}=IN{{R}_{p}}(c,{{f}_{\phi}})    \label{eq6}
\end{equation}

In the first step, as described in Eq. \eqref{eq4}, we parameterize the image $m$, where ${{f}_{\theta}}$ represents the initial function before parameterization, ${{f}_{\theta }^{m}}$ represents the parameterized secret function, and $IN{{R}_{p}}$ denotes the process of implicit representation. In the second step, as in Eq. \eqref{eq5}, the sender hides ${{f}_{\theta }^{m}}$ into another function ${{f}_{\phi}}$. We introduce a key $k$ to build the stego function ${{f}_{\phi}}$ based on ${{f}_{\theta }^{m}}$. In this paper, the key $k$ is used to control the function structure of ${{f}_{\phi}}$ and the location of ${{f}_{\theta }^{m}}$. In the third step, as in Eq. \eqref{eq6}, we utilize ${{f}_{\phi}}$ to achieve the implicit representation of a specific cover image $c$, obtaining the final stego function ${{f}_{\phi }^{c}}$, where $\phi$ is the parameter set of the stego function ${{f}_{\phi }^{c}}$.

For the message receiver, it is only necessary to use the secret key $k$ and the secret function ${{f}}_{\phi}^{c}$ obtained from the public channel to recover the secret message $m$. The receiver first recovers the secret function ${{f}}_{\theta}^{m}$ using Eq. \eqref{eq7}, and then utilizes the sampling process of INR to recover the secret message $m$ as follows:
\begin{equation}
    {{f}}_{\theta}^{m}=Recover({{f}}_{\phi}^{c},k)   \label{eq7}
\end{equation}
\begin{equation}
    m=INR_{s}({{f}}_{\theta}^{m})    \label{eq8}
\end{equation}

where $Recover$ represents the recovery process of the secret function ${{f}}_{\theta}^{m}$, and $INR_{s}$ represents the sampling process to recover the secret data $m$ from ${{f}}_{\theta}^{m}$.

\subsection{Data Representation}

From the principles of implicit representation, it can be seen that in order to train the parameters of this function, the data must be transformed into a set of coordinates and features, i.e., $m=\{({{x}}_{i},{{y}}_{i})\}_{i=1}^{n}$. Such a set of coordinates and features corresponds to the input/output pairs of the function, allowing us to learn the function's parameters. An individual sample point corresponds to a set of coordinates and features (i,e., an image is a set of position coordinates and individual pixel values). The use of a set representation of coordinates and features is very flexible, as it is independent of whether the data comes from a grid or the sampling resolution. The key is that constructing the problem entirely based on sets allows us to divide individual data points into subsets and train on them. Specifically, given an individual data point $s=\{({{x}}_{i},{{y}}_{i})\}_{i=1}^{n}$, such as a set of n pixels, we can randomly sample k elements(i.e., select $m$ pixels from the $n$ pixels of the entire image).

 Training on these subsets eliminates the direct dependency on data resolution. By adopting this approach, it is easy to obtain an implicit representation network that represents low-resolution data when higher resolution data is not required. Similarly, when training 3D shapes, there is no need to pass the entire voxel grid to the model. Instead, training can be done on subsets of the voxel grid, saving a significant amount of memory. This is impossible in standard convolutional models, as they are directly associated with the grid resolution. Furthermore, the training method based on the set representation of coordinates and features allows us to model more peculiar data, such as function distributions on manifolds. In fact, as long as we can define a coordinate system on the manifold (i.e., polar coordinates on a sphere), our steganographic method can be applied to all forms of data, whether it is used as secret messages or cover data.

\section{The Proposed Method}

\subsection{Secret Function}

In order to keep our method as simple as possible, we adopt MLP as the underlying network for implicit representation. The weight between the $l$-th and $(l+1)$-th layers can be written in matrix form as $W_{mn}^l$, and the bias can be written in vector form as the bias vector $b$. The dimensions of the weight matrix are $m \times n$, where $m$ represents the number of neurons on the right side of the two layers, and $n$ represents the number of neurons on the left side. The $i$-th row and $j$-th column of the matrix represent the weight between the $i$-th neuron in the right layer and the $j$-th neuron in the left layer. Thus, the equation can be simplified as described in Eq. \eqref{eq9} and Eq. \eqref{eq10}.
\begin{equation}
    {{s}_{m\times 1}^{l+1}}={{W}_{m\times n}^{l}}{{a}_{n\times 1}^{l}}+{{b}_{m\times 1}^{l}} \label{eq9}
\end{equation}
\begin{equation}
    {{a}_{n}^{l}}=\psi ({{s}_{n}^{l}}) \label{eq10}
\end{equation}

where $\psi (\cdot)$ represents the activation function. The parameters of the entire MLP network can be represented as a set of matrices and vectors, $P=\{{{W}_{{{m}_{l}}{{n}_{l}}}^{l}},{{b}_{{{m}_{l}}}^{l}}, l=1,...,L\}$. The network structure can also be represented as a vector consisting of all layer nodes, $S=\{{{n}_{1}^{in}},{{n}_{2}^{h}},{{n}_{3}^{h}},...{{n}_{L-1}^{h}},{{n}_{L}^{out}}\}$, where ${{n}_{1}^{in}}$ represents the dimension of the input layer and ${{n}_{L}^{out}}$ represents the dimension of the output layer. ${{n}_{l}^{h}}s $ represent the number of hidden neurons at layer $l$. We denote the randomly initialized MLP network that does not parameterize the secret message as ${{f}_{\theta}}=\{P,S\}$.

For a given secret message $m$, we use Eq. \eqref{eq1} to learn the parameters of the ${{f}_{\theta}}$. The resulting function representation is denoted as ${{f}_{\theta}^{m}}=\{{{P}_{\theta}},{{S}_{\theta}}\}$, referred to as the secret function.

\begin{figure}[htbp]\centering
	\includegraphics[width=0.8\linewidth]{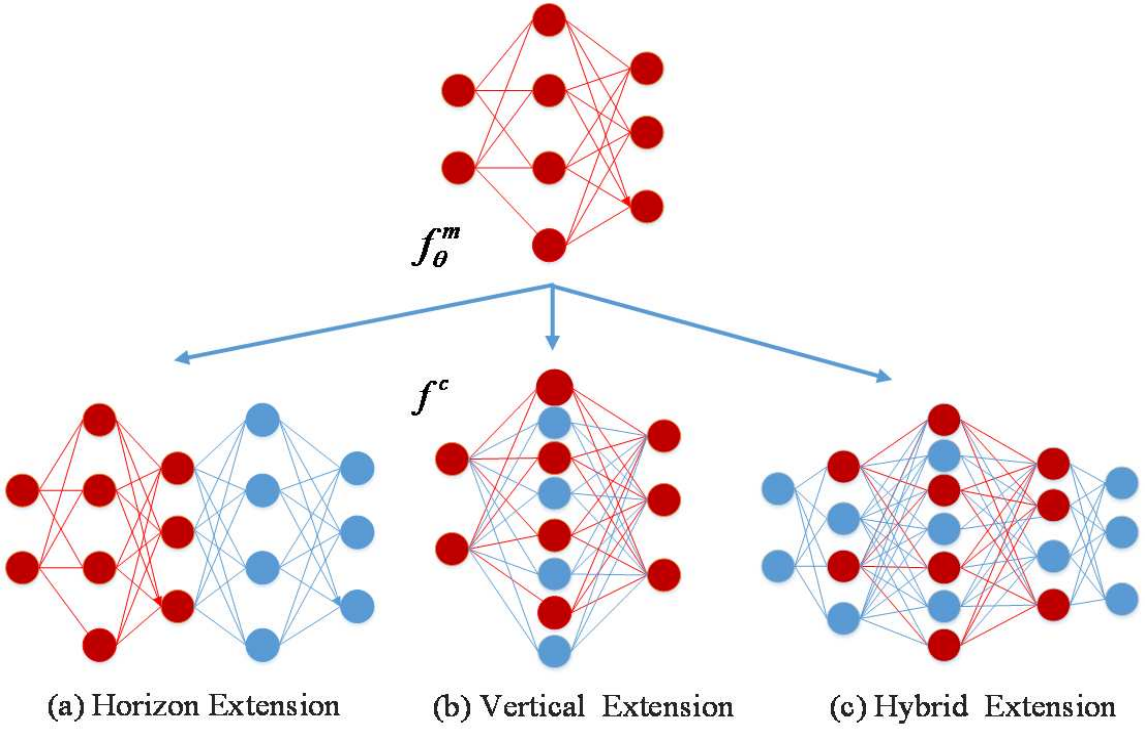}
	\caption{Three Strategies for Constructing Stego Functions}
    \label{FIG_3}
\end{figure}

\subsection{Stego Function}

In this paper, when considering the secret function as the secret message in traditional steganography, we can borrow the strategy of cover synthesis to directly construct the stego function using the secret function. In this paper, we restrict the cover synthesis to a special operation: function expansion. Specifically, the stego function is constructed by adding neurons to the secret function while keeping its structure and parameters unchanged. Since the implicit representation can adopt relatively simple network structures such as MLP, we can create a new stego function by inserting a large number of new neurons without affecting performance. Another reason for choosing expansion instead of selection, as in \cite{SteganographyofSteganographicNetworks}, is that after implicit representation, the neural network has already overfit the data, and removing any neuron may result in the secret function being unable to sample the secret data.

More importantly, we want the expanded stego function to have certain functionality, making it more concealable during transmission (i.e., using some behavior to conceal the steganographic behavior). Typically, we have two choices. One is to use this stego function (network) to perform a regular machine learning tasks, such as image classification, as done in \cite{SteganographyofSteganographicNetworks,TowardsDeepNetworkSteganography}. In this paper, we adopt another method that is consistent with the secret message, namely using the implicit representation to conceal the steganographic behavior. When both the secret function and the stego function are used to represent the same type of data, taking images as an example, we design the following three strategies for function expansion, as shown in Fig. \ref{FIG_3}.

(1)\textit{ Horizontal Expansion}. This strategy refers to inserting new layers after the secret function to construct a new function, while keeping the previous network parameters and structure unchanged, as shown in Fig. \ref{FIG_3} (a).

(2) \textit{Vertical Expansion}. Another method of expanding the function is to keep the number of layers unchanged. Since both the cover function and the secret function are used to represent images, we only need to increase the number of hidden units in the hidden layers, except for the input layer and the final output layer, as shown in  Fig. \ref{FIG_3} (b).

(3) \textit{Mixed Expansion}. Mixed expansion involves both vertical and horizontal expansions, i.e., increasing the number of layers and the number of neurons in existing hidden layers at the same time. This is a general operation, as shown in Fig. \ref{FIG_3} (c), where all the content, parameters, and structure of the secret network are hidden within a single network.

In fact, vertical expansion and horizontal expansion can both be seen as special cases of mixed expansion operations. We aim to use simple expansion forms as much as possible to achieve our objectives.

When the secret function and the cover function are used to represent different types of data, we can naturally adopt the mixed expansion for stego function construction. For example, hiding the implicit representation of an image within the implicit representation of a 3D model. Once the structure of the stego function $f_{\phi}$ is determined, we can train it to represent a specific cover data $c$. It should be noted that the common characteristic of these three expansion methods is that the secret function $f^m_{\theta}$ remains intact within the stego function, and the dependency between layers is not disrupted. This will assist us in designing the recovery of the secret function.

Since the structure of the cover function needs to be determined before expansion, we can consider the function expansion operation as relying on the shared secret information $k$, as shown in Eq. \eqref{eq5}. In our scheme, the key $k = \{ k_1, k_2, ..., k_L\}$ indicates the positions of $f^m_{\theta}$ within $f_{\phi}$, where each layer corresponds to a binary stream $k_l$ of $d_{k_{l}}$ length, representing the number of neurons in that layer. Each bit corresponds to a neuron, where 0 represents the position of the masked neuron, and 1 represents the position of the secret neuron. For example, the key $k$ corresponding to the stego function $f_{\phi}$ in Fig. \ref{FIG_3} (c) is $\{{00, 1010, 1010101, 1101, 000\}}$.

From the perspective of constructing $f_{\phi}$, the cover function is obtained by expanding under the influence of the key $k$, with some neuron positions and parameters fixed, and using random numbers. This method of constructing a function containing secrets based on the key $k$ is similar to the Cardan grille \textsuperscript{\cite{GenerativeSteganographybySampling}}. The key $k$ itself serves as a Cardan grille that marks the positions of the hidden secret units. In this paper, we migrate the Cardan grille method from grid data to the function domain. Therefore, our method can also be called Cardan Grille-based Stego Function Generation (CG-SFG).

\subsection{Training of the Stego Function}
We use \eqref{eq1} to achieve the implicit representation of the cover image $c$ by the stego function $f_{\phi}^c$. $\phi$ includes two different types of parameters: the original parameters $\theta$ of the secret function $f_{\theta}^{m}$, and the parameters $\phi_e$ of the expanded neurons. To ensure that we can extract $f_{\theta}^{m}$ from $f_{\phi}^c$ without loss, during the training process, we fix the parameters $\theta$ of $f_{\theta}^{m}$ and only adjust the parameters $\phi_e$ of the expanded neurons. To do this, we introduce a mask $M^k$ to partially optimize $f_{\phi}$. $M^k$ is a binary mask vector $v^k$ with the same size as the parameter set $\phi$ of $f_{\phi}$, as computed in Eq. \eqref{eq14}
\begin{equation}
v^k[i]=
\begin{cases}
1 \text{    if $\phi[i] \in \phi_e$}\\
0 \text{    if $\phi[i] \in \theta$}
\end{cases} \label{eq14} 
\end{equation} 
where $\phi[i]$ is the $i$-th parameter of $f_{\phi}^{c}$, and the number of parameters of $f_\phi$ can be calculated from $P^c$ or $S^c$. The value of the vector $v^k$ can be computed from the key $k$, which we will explain in detail in Section \textit{Secret Function Recovery}.

Let $\eta$ be the learning rate and $\odot $ denotes element-wise multiplication. We update $f_{\phi}^{c}$ using gradient descent as described in Eq. \eqref{eq15}:

\begin{equation}
    \phi = \phi - v^k \odot  \eta \nabla _\phi L_c    \label{eq15}
\end{equation}

Through training, we obtain the parameterized representation of the stego function $f_{\phi}^{c}$ for the cover image c.

\subsection{Secret Function Recovery}
The stego function $f_{\phi}^c$ can be transmitted over a public channel, and regular users can only sample the cover data $c$ represented by the stego function. Since the secret function $f_{\theta}^m$ remains unchanged during the construction and training of the stego function, the receiver of the secret message can recover the secret function $f_{\theta}^m$ without loss after obtaining the function $f_{\phi}^c$ using the key $k$. By using the key $k$, we can not only determine which are the hidden neurons, but also obtain the weight information between those neurons. That is, we can obtain the parameters $P_{\theta}^m$ and the structure $S_{\theta}^c$ of the secret function $f_{\theta}^m$ from the key $k$.

\begin{figure}[htbp]\centering
	\includegraphics[width=\linewidth]{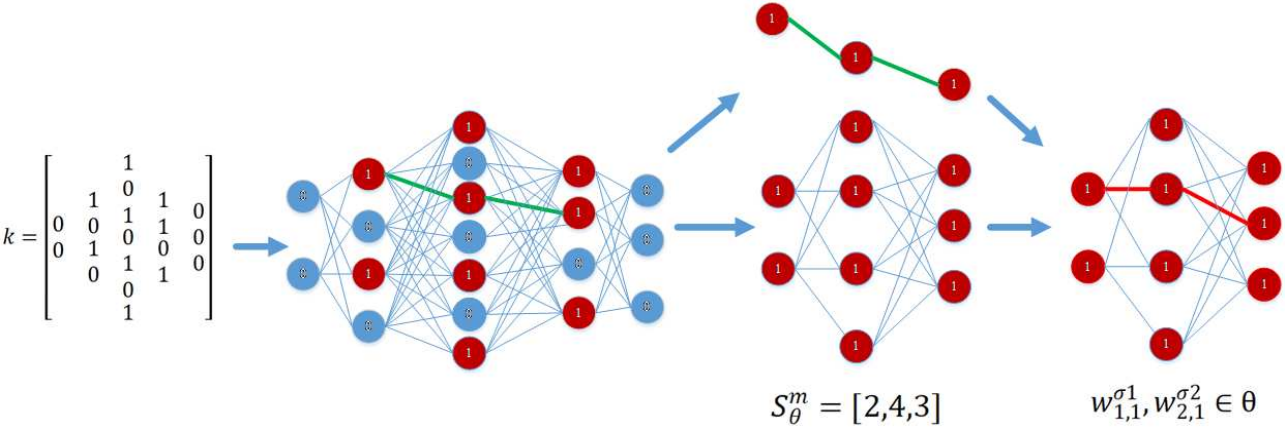}
	\caption{Secret Function Recovery.}
    \label{FIG_5}
\end{figure}

As shown in Fig. \ref{FIG_5}, the receiver first overlays the values of $k$ onto the neuron nodes $n$ of the stego function $f_{\phi}^c$, (i.e., if $k^l_{i_{l}}= 1$, then $n(i_l)^l = 1$, where $n(i_l)^l$ represents the $i_l$-th node on the $l$-th layer). We only need to keep the weights ${{w}_{{{i}_{l-1}},{{i}_{l}}}^{l-1,l}}$ that satisfy both $n(i_l)^l = 1$  and $n(i_{l-1})^{l-1} = 1$, and then we can obtain the weight matrix of layer $l$ in $f_{\theta}^m$ as $W^{\theta}_{l, m \times n}$, where $m$ and $n$ represent the dimensions of the weight matrix. Thus, the parameters of the secret function $f_{\theta}^m$ can be represented as described in Eq. \eqref{eq16}.

\begin{equation}
    {{P}_{\theta }^{m}}=\{{{W}_{l}^{\theta }}=\{{{w}_{{{i}_{l-1}},{{i}_{l}}}^{l-1,l}}{{\}}_{m\times n}}{{\}}_{i=1:{{L}_{c}}}} \label{eq16}
\end{equation}

Finally, by counting the number of 1s in each layer of $k$, which represents the number of secret neurons, we can obtain the model structure as described in Eq. \eqref{eq17}

\begin{equation}
    S_{\theta}^{m}=\{count_{nonzero}(k_i)\}_{i=1:L_{c}} \label{eq17}
\end{equation} 

\subsection{Secret Message Extraction}
To extract the secret message $m$ from the recovered secret function $f^{\theta}_m$, it is essentially a problem of sampling from the implicit representation function, as described in Eq. \eqref{eq18}.

\begin{equation}
    \{{{y}_{i}}{{\}}_{i=1}^{n}}={{f}_{\theta }^{m}}(\{{{x}_{i}}{{\}}_{i=1}^{n}})   \label{eq18}
\end{equation}

We only need to convert ${\{({x_i}, {y_i})\}}_{i=1}^n$ to the original format of the corresponding secret data. Taking images as an example, we can input a set of coordinates ${\{({x_1}, {x_2})\}}_{i=1}^n$ to $f_{\theta}^m$ to obtain the corresponding RGB pixel values $\{{p_i}\}_{i=1}^n$.

\section{Experiment}
\subsection{Setup}
We evaluate our model on various resolution images from CelebAHQ\textsuperscript{\cite{Progressivegrowing}}, Div2k \textsuperscript{\cite{Div2k}} and COCO dataset \textsuperscript{\cite{coco}}, and climate data from the ERA5 dataset\textsuperscript{\cite{Peubeyand}}. We also use common datasets NeRF Synthetic\textsuperscript{\cite{Nerf}} with forward 360\textdegree scenes {lego, drums, chair} from NeRF-Synthetic. We use an MLP for secret function and stego function to represent the secret message and the cover data. Our model is implemented using PyTorch \textsuperscript{\cite{Pytorch}}, and we utilize the implicit representation framework provided in \textsuperscript{\cite{GenerativeModelsasDistributions}}. All training is conducted on a single 2080Ti GPU with 11GB memory.

Taking CelebAHQ as an example, the size of the secret message image is 64x64. We represent the secret function, which is the network structure for representing the secret image, as $[2, 64, 64, 64, 3]$. In this structure, the input layer dimension is 2, representing the pixel coordinates, and the output dimension is 3, representing RGB values. The number of neurons in the three hidden layers is all set to 64. Since the input and output layers depend on the data type, in the experiment, we use the hidden layers vector $[64, 64, 64]$ to represent the entire network structure. We train the model for 2000 epochs using the stochastic gradient descent algorithm with a learning rate of $\eta=0.0001$.

 The size of the cover image is also $64 \times 64$. We represent the cover image with cover function which structure is $[1024, 512, 1024]$. We train the model for 2000 epochs using the stochastic gradient descent algorithm with a learning rate of $\eta=0.001$.

\begin{figure}[h]\centering
	\includegraphics[width=\linewidth]{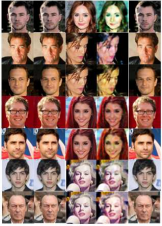}
	\caption{A group of experimental results on the CelebA-HQ database.}
    \label{FIG_6}
\end{figure}

In Fig. \ref{FIG_6}, each column represents the secret image, image sampled from the secret function, cover image, cover image sampled from the stego function, and the secret image  sampled from the stego function, respectively. From the above results, it can be seen that although the stego function introduces some distortion to the cover image, this may be due to fixed parameters leading to some loss in the network's fitting effect, but it can still capture the overall information of the cover image. Most importantly, with the key, the secret function can be recovered from the stego function, and the secret message obtained by sampling the secret function has a high similarity to the original secret message, which is almost indistinguishable.

\subsection{Performance}

\subsubsection{Representation of The Secret Message}
\begin{figure}[h]\centering
	\includegraphics[width=\linewidth]{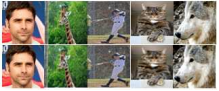}
	\caption{The images sampled by an implicit representation network at different resolutions.}
    \label{FIG_7}
\end{figure}

For the secret image $m$, we use the Peak Signal-to-Noise Ratio (PSNR) of the original image and the image obtained by sampling the secret function to measure the implicit representation capability. Fig. \ref{FIG_7} shows the representation results of images with different resolutions in the case of an MLP network structure of [64, 64, 64]. From left to right, the resolutions are 64x64, 128x128,  256x256, 512x512, and 1024x1024, respectively, in which the 64x64 image comes from CelebAHQ dataset, the 128x128 and 256x256 images come from COCO dataset, and the 512x512 and 1024x1024 images come from the Div2K dataset.

\begin{figure}[h]\centering
	\includegraphics[width=\linewidth]{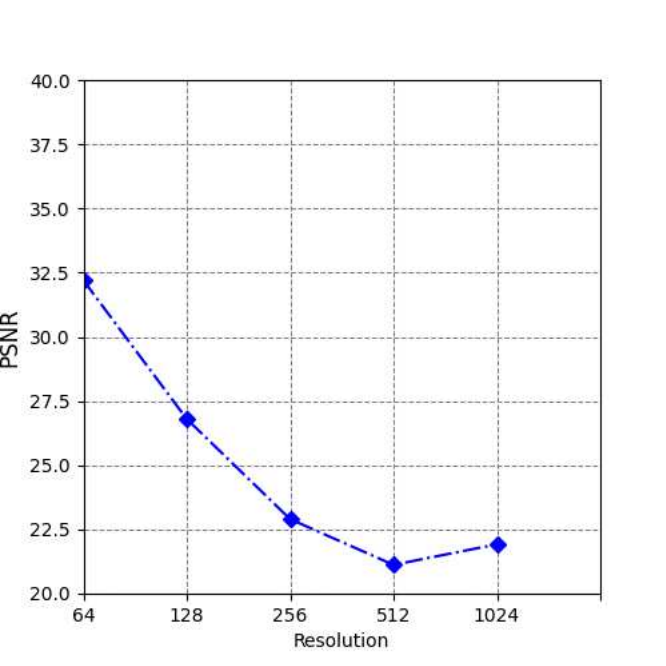}
	\caption{The PSNR curve of images obtained at different resolutions.}
    \label{FIG_8}
\end{figure}
It can be observed that a simple network can represent secret messages of multiple resolutions. In practice, although implicit representation cannot fully represent secret messages without any distortion, it is acceptable in image steganography since images typically contain a significant amount of redundant information. Under the low embedding capacity constraint in traditional steganography, there is a need for high accuracy in message extraction. Fortunately, with an increase in embedding capacity, particularly in steganography techniques using images as secret messages, the focus is mostly on preserving the integrity of semantic information, and some distortion to the original image is acceptable in many cases. Therefore, we adopt image quality evaluation standards such as PSNR to validate our algorithm. 

Figure \ref{FIG_8} shows the PSNR values of the secret image and the sampled secret image from the secret function at different resolutions. It can be seen from the figure that as the resolution increases, the PSNR decreases to some extent. This is because the method of calculating differences only treats the image as isolated pixels and ignores some of the visual features contained in the image contents.

\subsubsection{Recoverability}

\begin{table}[h]
\caption{The BER for different structures of secret functions}\label{tab1}%
\begin{tabular}{@{}llll@{}}
\toprule
secret function $f^{m}_{\theta}$ & stego function $f^{c}_{\phi}$ & extracted  $f^{m}_{\theta_{e}}$ & BER \\
\midrule
 64,64,64 & 128,128,128 & 64,64,64 & 0.0\\
 256,128,256 & 512,256,512 & 256,128,256 & 0.0 \\
 512,256,512 & 1024,512,512 & 512,256,512 & 0.0 \\
 256,128,128,256 & 512,256,256,512 & 256,128,128,256 & 0.0\\
\bottomrule
\end{tabular}
\end{table}
We calculate the bit error rate (BER) between the original secret function ${{f} _ {{\theta} }^{m}}$ and the secret function ${{f} _ {{{\theta} _ {e}}}^{m}}$ extracted from the stego function ${{f} _ {\phi} ^ {c}}$. Table. \ref{tab1} presents the error rates of different architectures of secret function models in extracting secret models from the stego function. It can be seen that due to fixing the parameters of the secret function in the stego function, the secret function can be extracted without any loss, and the performance of the images represented by the secret function is not affected.

\subsubsection{Fidelity}

To evaluate the fidelity of the stego function ${{f} _ {\phi} ^ {c}}$, we trained a clean model ${{f} _ {\gamma} ^ {c}}$ with the same architecture as $ {{f} _ {\phi}^{c}}$. The training of ${{f} _ {\phi} ^ {c}}$ and ${{f} _ {\gamma} ^ {c}}$ was conducted on the same training set, and all parameters in ${{f} _ {\gamma} ^ {c}} $were adjusted for optimal performance. Additionally, we used the same test set to evaluate the performance of these two models. 
\begin{figure}[h]\centering
	\includegraphics[width=\linewidth]{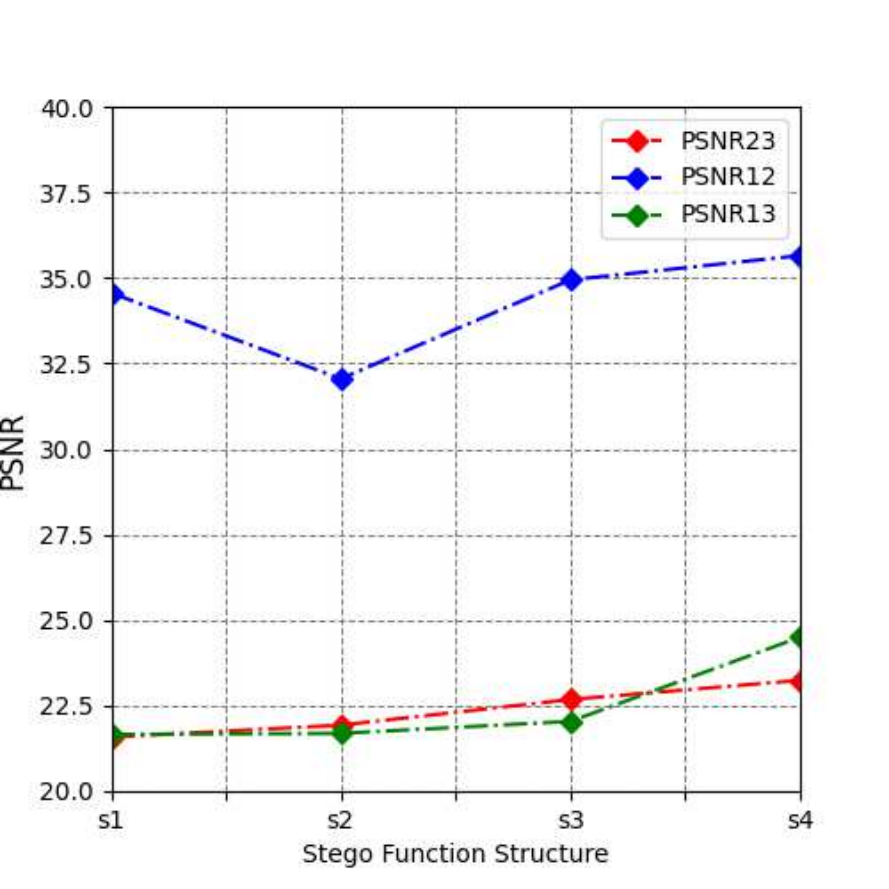}
	\caption{The PSNR curves corresponding to the original carrier, the sampled image of the clean function, and the sampled image of the stego function, respectively. }
    \label{FIG_9}
\end{figure}

We have used four network structures to evaluate the performance of the stego function, as shown in the second column of Table \ref{tab1}. We also calculated the sampled results of the original cover image on both the clean model and the stego model, and used PSNR to measure their representation capabilities. 
\begin{figure}[h]\centering
	\includegraphics[width=\linewidth]{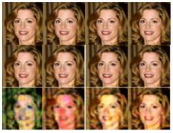}
	\caption{ The original carrier, the sampled image of the clean function, and the sampled image of the stego function.}
    \label{FIG_10}
\end{figure}
The PSNR curves are shown in Fig. \ref{FIG_9}, where PSNR12 represents the PSNR between the original image and the sampled image from the clean model, PSNR13 represents the PSNR between the original image and the sampled image from the stego function, and PSNR23 represents the PSNR between the sampled images from the clean model and the stego function, respectively. It can be observed that fixing some parameters indeed leads to a decrease in the representation capability of the stego function. As the parameters of the stego function model increase, the proportion of parameters of the secret function in the stego function decreases, resulting in an improvement in the representation capability of the stego function. 
Fig. \ref{FIG_10} presents the sampled results of the same cover image using different structures of clean functions and stego functions for representation. The first row represents the original cover image. The second row shows the sampled image obtained from the clean function, while the third row shows the sampled image obtained from the stego function.
\begin{figure}[htbp]\centering
	\includegraphics[width=\linewidth]{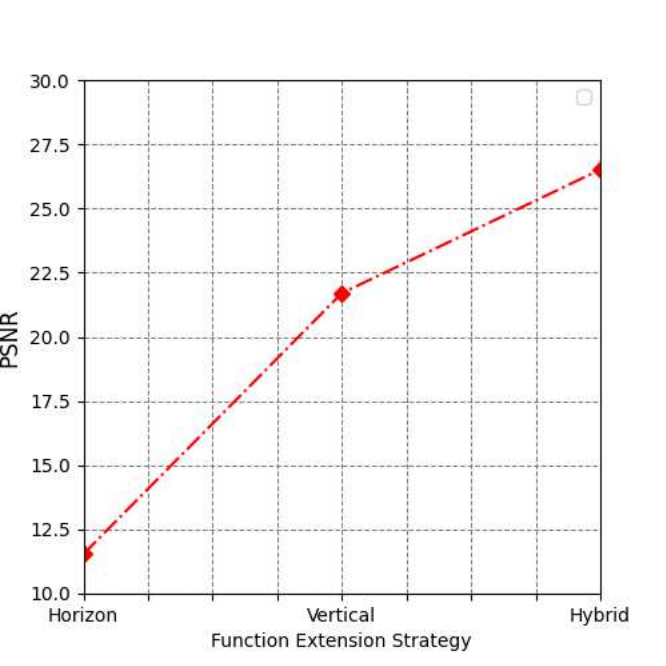}
	\caption{PSNR curves under different expansion strategies.}
    \label{FIG_11}
\end{figure}
\begin{figure}[h]\centering
	\includegraphics[width=\linewidth]{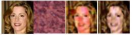}
	\caption{The images sampled from stego functions obtained by different expansion strategies.}
    \label{FIG_12}
\end{figure}
We also evaluated the three different stego function strategies mentioned to verify the effectiveness of various function extension strategies. The three expansion strategies of the stego function are as follows: [64,64,64, 256,256,256],[256,256,256] and [256,256,256,256]. As shown in Fig. \ref{FIG_11}, the stego function generated by the the hybrid mode exhibits the best performance. Figure \ref{FIG_12} show the sampled result under different modes. The first column represents the original cover image, and the second to fourth columns show the images obtained by sampling from the stego function constructed using horizontal, vertical, and hybrid modes, respectively.

\subsubsection{Capacity}

\begin{figure}[h]\centering
	\includegraphics[width=\linewidth]{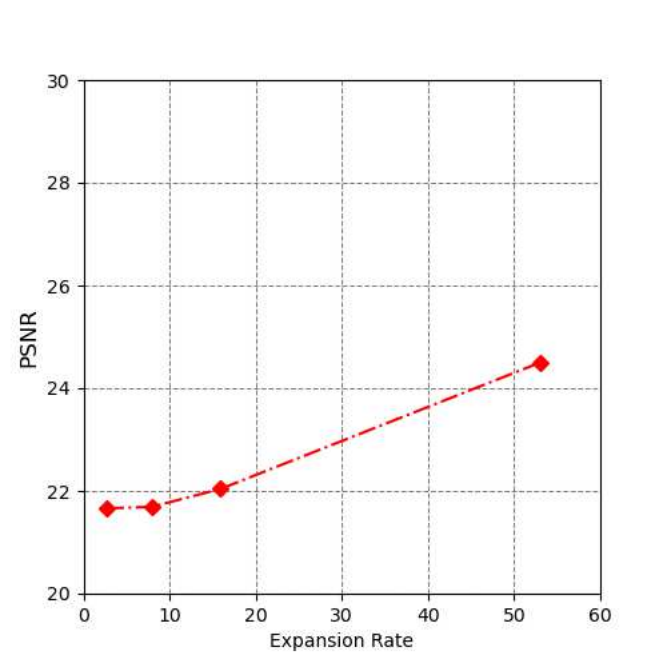}
	\caption{The PSNR between the sampled images from the stego function and the original cover image under different expansion rates.}
    \label{FIG_13}
\end{figure}

Our approach hides the parameters of one function within another function, and the capacity of the proposed method can be evaluated based on the expansion rate of the parameters. The function expansion rate is defined as described in Eq. \eqref{eq19}

\begin{equation}
    e = N_{stego}/N_{secret}\label{eq19}
\end{equation}
where $N_{stego}$ and $N_{secret}$ represent the number of parameters in the stego function and secret function, respectively. Thus, we have $e \geqslant 1$. We further evaluate how the performance of our proposed method changes as the expansion rate varies. Fig \ref{FIG_13} displays the implicit expressive ability of the stego function at different expansion rates.The structure of the secret function is [64, 64, 64], and the structure of the masking function is given in the second column of Table \ref{tab1}.
As can be seen, as the expansion rate increases, the stego model gradually improves its representation ability on the cover.
\subsubsection{Diversity of Data Representations}

\begin{figure}[h]\centering
	\includegraphics[width=\linewidth]{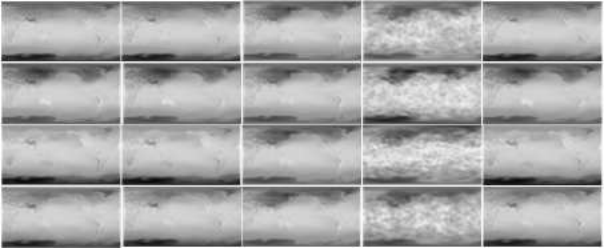}

	\caption{The result on climate data.}
    \label{FIG_14}
\end{figure}

Implicit representation can be used to represent various types of data.  We also conducted experiments on climate data from the ERA5 dataset\textsuperscript{\cite{Peubeyand}}. The formats of ERA5 meteorological data are generally Grid and NetCDF formats. We convert Grib format data into \textit{npz} data formats. In order to display the data intuitively, we save data as image format, as shown in the Fig. \ref{FIG_14}. We use the data from January, April, July, October of 1979 as the secret message, and the temperature data from December of 2020 as the cover data. The structure of the secret function is [128, 128, 128], and the structure of the stego function is [512, 512, 512, 512]. We train all the models for 1000 epochs with $\eta = 0.001$. From left to right, each column represents: original secret message data, secret message data sampled from the secret function, original cover data, sampled cover data from the stego function, and recovered secret message data from the stego function. From top to bottom, each row represents the experimental results using secret data from different months. The results show that with implicit representation, we can easily embed different types of message data into the stego function. 

\begin{figure}[htbp]\centering
	\includegraphics[width=\linewidth]{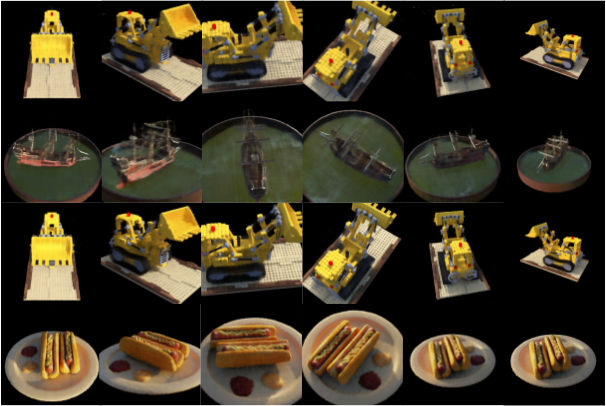}

	\caption{The result on NeRF Synthetic 3D data.}
    \label{FIG_15}
\end{figure}

On NeRF-Synthetic, we adopt \cite{chenmingxiang110} as the NeRF backbone architecture for efficiency. This network consists of four blocks, each block is composed of some linear layers and non-linear activation functions. The first stage of training for secret 3D message is performed according to the standard recipes of this implementation. We then extend the network that represents secret messages by maintaining four blocks and increasing the number of fully connected nodes in each block. The network that masks the carrier while keeping the original parameters unchanged is retrained. In order to display the 3D data, we save data as image format, as shown in the Fig. \ref{FIG_15}. Here we present two sets of experimental results, from top to bottom, representing stego 3D data using LEGO and representing secret 3D data using hot dogs, as well as stego 3D data using LEGO and representing secret data using 3D boat data. From the results, it can be observed that our method also performs effectively in representing implicit representations of 3D data. Similar to image representation, in the representation of disguised data, namely LEGO data, the rendering quality is somewhat inferior compared to the original LEGO data rendering quality.
\begin{table}
\caption{The Undetectability of the proposed method}\label{tabn}

\begin{tabular}{@{}ccccc@{} }
\toprule
\makecell{Detection \\method}   & \makecell{Accuracy \\(Image)}  & \makecell{Accuracy \\(Climate Image)} & \makecell{Accuracy \\(3D)}  & Mean \\
\midrule
SVM-poly    & 0.53   & 0.52  & 0.53  & 0.53\\
SVM-rbf    & 0.55   & 0.54  & 0.54  & 0.54\\
SVM-sigmoid    & 0.54   & 0.55  & 0.53 & 0.54 \\
\bottomrule
\end{tabular}

\end{table}
\begin{table*}[htbp]
\caption{Comparison with other Steganography Methods}\label{tab3}
\resizebox{\textwidth}{!}{%
\begin{tabular}{@{}llllll@{}}
\toprule
Category & Method  & Capacity (bpp/er) & BER/PSNR & Eoob/Detection Rate & PSNR/ACC \\
\midrule
Traditional Steganography & \cite{liao2022adaptive} & 0.50/- & 0.0/- & 33.3/- & -/- \\
Traditional Steganography & \cite{Automaticsteganographicdistortionlearning} & 0.40/-& 0.0/-  & 16.20/- & -/-\\
Traditional Steganography & \cite{tang2020automatic} &0.50/- & 0.0/- & 26.49/- & -/-\\
Deep Steganography & \cite{zhu2018hidden} & 8/- & -/35.70  & -/76.49 & 30.09-62.12/-\\
Deep Steganography & \cite{baluja2020hiding} & 8/- & -/34.13  & -/99.67 & 32.0-41.2/-\\
Deep Steganography &\cite{jing2021hinet} & 8/- & -/46.78  & -/55.86 & 44.60-52.86/-\\
Model Steganography& \cite{SteganographyofSteganographicNetworks} & -/1.52 & 0/-  & -/52.50(SVM) & -/0.93 \\
Model Steganography & \cite{TowardsDeepNetworkSteganography} & -/0-1.65 & 0.00668/-  & -/47.5(SVM) & -/0.938 \\
StegaINR & Ours & 24-192/2.66-53.04 & 0/-  & -/53.0(SVM) & 24.5/- \\
\bottomrule
\end{tabular}%
}
\end{table*}

\subsubsection{Undetectability}

In our method, the undetectability of the stego function (model) is a crucial concern. We establish a model pool that consists of multiple stego functions and their structurally identical clean functions. Both the stego and clean functions are trained to implicitly represent the same cover (Image,Climate Image,3D data). Since our method only requires training on a single cover, we can create more nature and stego models. Ultimately, we construct a dataset consisting of 1200 stego functions and 1200 clean functions, each trained on 1200 different pairs of cover data,.

Training a classifier to detect differences in parameter distributions between stego functions and clean functions is a direct approach. Following this idea, we randomly select 1000 parameters from each model as statistical measures of the parameters, resulting in a 1000-dimensional feature vector used to train a Support Vector Machine (SVM) classifier \textsuperscript{\cite{LIBSVM}}. Specifically, we randomly select 1000 pairs of features from stego and clean functions for training, while the remaining 200 pairs are used for testing. The results in Table \ref{tabn} indicate that the SVM method cannot detect the presence of secret functions from stego functions, with a detection mean accuracy close to $50\%$, where the SVM-poly refers to SVM with a polynomial kernel, SVM-rbf refers to SVM with a Gaussian (RBF) kernel, while SVM-simoid refers to SVM with a sigmoid kernel.

\subsection{Comparison}
As shown in  Table \ref{tab3}, we compared our proposed method with existing state-of-the-art multimedia (image) steganography schemes and model steganography schemes based on deep neural networks, evaluating them in terms of capacity, recoverability, detectability, and fidelity (only for model watermarking schemes). To ensure a fair comparison among different types of information hiding schemes, for multimedia steganography, we assume that the data to be hidden is a secret image of size $64 \times 64$. For model steganography, we assume that the data is an MLP with a hidden layer size of [64,64,64].

For traditional steganography schemes\textsuperscript{\cite{liao2022adaptive,Automaticsteganographicdistortionlearning,tang2020automatic}} we copied the reported End-Of-Out-Band (EOOB) rate from each paper under a specific payload, where the payload refers to the number of bits per pixel (i.e., how many bits can be embedded per pixel). EOOB is calculated as the average of the false positive rate and the false negative rate. 

For deep steganography schemes \textsuperscript{\cite{zhu2018hidden,baluja2020hiding,jing2021hinet}}, we copied the results reported in Zhou\textsuperscript{\cite{Deepresidualnetworkforsteganalysisofdigitalimages}} for Perceptual Signal-to-Noise Ratio (PSNR) at 8bpp payload (on ImageNet) and detection accuracy.

It can be observed that, compared to traditional multimedia steganography methods, our approach does not have an advantage in terms of image quality. Moreover, what is interesting is that when our method is regarded as a multimedia steganography scheme. From the perspective of transforming the original message image into the cover image, we can represent messages of arbitrary resolutions as continuous functions, thereby achieving a high-capacity steganography scheme. This is mainly due to representing data as functions, which makes the size of the data independent of the resolution. In terms of detection rate, we have similar characteristics to model-based steganography.

\section{Discussion}

\textit{Limitations}:  While our method can be regarded as a fusion of multimedia steganography and network model steganography, it does not outperform state-of-the-art image or model steganography in terms of performance. In designing our method, we aimed for simplicity by directly utilizing function extensions to construct stego functions. However, the approach of fixing the parameters of the secret function results in reduced performance when representing the cover data. Although increasing the expansion rate can improve representational performance to some extent, the increased training time and significant rise in parameters may raise concerns about this approach.

Another fundamental issue is that, despite the considerable representational capability of implicit representation in theory, there remains a gap between the implicitly represented images and other data and the original data. However, due to the significant redundancy present in multimedia data like images, even if there is such a gap, steganography is generally considered successful as long as the semantic differences are small. However, in scenarios where the implicitly represented data must be completely identical to the original data, our approach is difficult to apply. For example, in the case of hiding text data, even if the implicit neural network overfits, it cannot completely and accurately represent text data due to its small redundancy and low tolerance for error. Even minor modifications can lead to semantic changes, making this method unsuitable for scenarios requiring lossless transmission of secret messages

\textit{Future work}: In this paper, we have presented a pioneering approach to steganography by leveraging implicit representation, thereby offering a fresh implementation strategy for multimedia steganography. With the emergence of implicit representation techniques like NeRF, we believe that our method can be further enhanced to cater to evolving requirements. Notably, our approach of representing data as functions or solely relying on coordinates and features opens up steganography applications in diverse fields, not limited to secure communication but also encompassing embedding operations for data storage purposes. This marks the initial stride we have taken in this direction. While algorithmic refinements are crucial, our immediate next step is to devise superior techniques for constructing stego functions that yield improved representational performance for cover data. Furthermore, we intend to evaluate our proposed solution on a wider range of datasets, including 3D data and more. Additionally, a key focus of future research will be the assessment of undetectability between stego functions and clean functions.

\section{Conclusion}
This paper introduces StegaINR, which is the first automatic framework for steganographic information embedding in implicit neural representations. Our stego function acts as both the decoder and stego carrier, allowing for secure transmission through public channels. By treating the secret message data as a continuous function, our method can be applied to various types of information. To our knowledge, this is the first time implicit representation has been introduced to the field of steganography, and our experiments show that the approach can achieve secure and high-capacity steganography in implicit representations of images and weather data. We outline the relevant components necessary to achieve this functionality, along with potential challenges and insights for future developers.

\vskip 2mm
\noindent
\textbf{Acknowledgment}
\vskip 2mm
\noindent
 This work was supported in part by the National Key R\&D Program of China under grant 2021YFB3100901, NSF of China under Grant 62074131, 62272478 and 62202496, and Shaanxi Provincial Key R\&D Program 2023-ZDLGY-32.

\renewcommand\refname{\large

\begin{strip}
\end{strip}

\mbox{}
\clearpage
\clearpage

  \end{document}